\documentclass[aps,prl,twocolumn,superscriptaddress,showpacs]{revtex4}

\usepackage{amssymb}
\usepackage{amsmath}
\usepackage{graphicx}
\usepackage{natbib}

\begin{document}

\title{Quantum Coherence in a One-Electron Semiconductor Charge Qubit}

\author{K. D. Petersson}
\author{J. R. Petta}
\affiliation{Department of Physics, Princeton University, Princeton, NJ 08544, USA.}
\author{H. Lu}
\author{A. C. Gossard}
\affiliation{Materials Department, University of California at Santa Barbara, Santa Barbara, CA 93106, USA.}

\date{\today}

\begin{abstract}
We study quantum coherence in a semiconductor charge qubit formed from a GaAs double quantum dot containing a single electron. Voltage pulses are applied to depletion gates to drive qubit rotations and non-invasive state readout is achieved using a quantum point contact charge detector. We measure a maximum coherence time of $\sim 7$ ns at the charge degeneracy point, where the qubit level splitting is first-order-insensitive to gate voltage fluctuations. We compare measurements of the coherence time as a function of detuning with numerical simulations and predictions from a $1/f$ noise model.
\end{abstract}


\maketitle

The key requirement that a quantum computer be scalable has motivated recent work exploring coherent control of two-level systems in the solid state. A large effort has focused on quantum dots, where quantum control of both single spin and two spin ``singlet-triplet" qubits has been demonstrated \cite{koppens06, petta05, foletti09, petta10}. While progress has been rapid, reliable two-qubit gates are required in order to scale to larger system sizes \cite{Hanson_RMP}. Proposals for two-qubit gates rely on a charge-noise-susceptible exchange interaction \cite{loss98,petta05,taylor05,coish,Hu_PRL_2006}. Developing a quantitative understanding of the charge noise environment, and how it impacts quantum coherence, is therefore crucial for quantum dot approaches to quantum information processing.

Early demonstrations of quantum coherence in the solid-state took place using charge qubits, which can have $\sim$ 100 ps gate operation times and relatively long coherence times \cite{Nakamura_Nature_1999,vion02}. Nakamura \textit{et al.} demonstrated charge coherence in a superconducting Cooper pair box (CPB), where the state of the qubit is determined by the number of Cooper pairs on a superconducting island \cite{Nakamura_Nature_1999, Astafiev_PRL_2004}. In semiconductor systems, a charge qubit can be formed by isolating an electron in a tunnel-coupled double quantum dot (DQD) \cite{hayashi03,fujisawa06}. Here the state of the qubit is set by the position of the electron in the double well potential. Coherent control of a GaAs charge qubit has been demonstrated \cite{hayashi03, fujisawa04}, along with correlated two qubit interactions \cite{shinkai09}. However, precise values of the coherence time are unknown in GaAs, since state readout in past experiments involved transport through the DQD with strong coupling to the leads, typically limiting coherence times to $\sim 1$ ns due to cotunnelling \cite{hayashi03}. In addition, each dot contained a few tens of electrons, potentially complicating the qubit level structure.

\begin{figure}[t]
\begin{center}
		\includegraphics[scale=1]{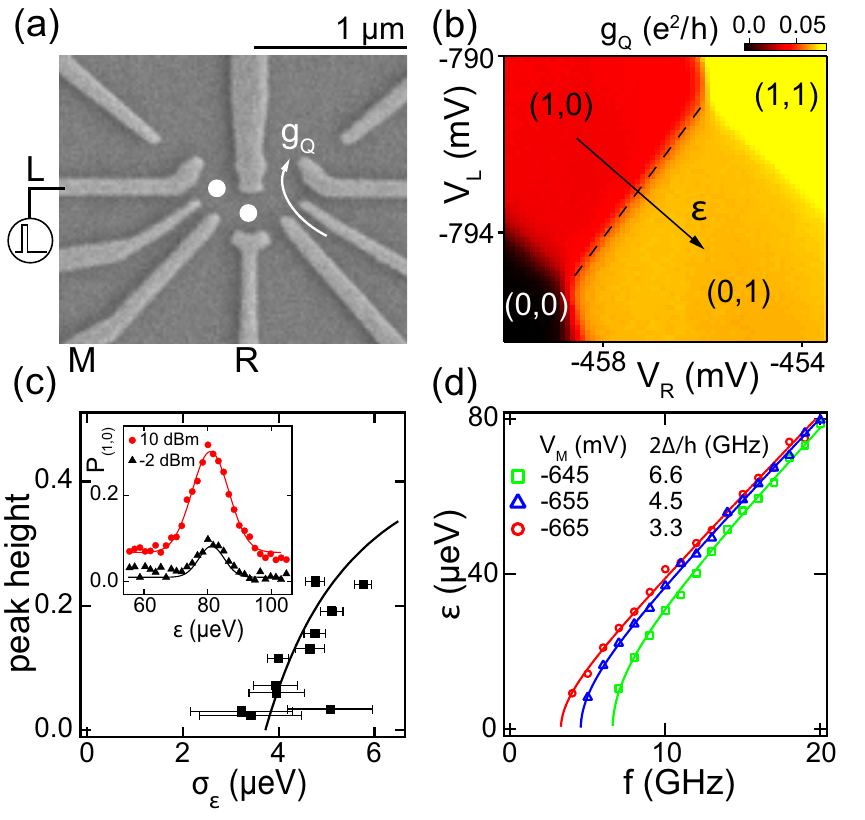}
\caption{\label{fig1} (Color online) (a) Scanning electron micrograph of a device similar to the one measured. (b) Charge sensor conductance, $g_Q$, measured near the $(1,0)-(0,1)$ charge transition. (c) Peak height versus linewidth of the microwave-induced resonances for different applied microwave powers. The solid curve is a fit to theory (see text). Inset: Left dot occupation, $P_{(1,0)}$, as a function of detuning, $\epsilon$, for a microwave peak at two different powers. The solid lines show Gaussian fits to the data. (d) Microwave spectroscopy data acquired for three different values of $V_M$.}
\end{center}	
\vspace{-1.0 cm}
\end{figure}

In this Letter, we demonstrate coherent control of a tunable GaAs charge qubit containing a single electron. In previous experiments, voltage pulses were applied to the drain contact of the DQD for quantum control \cite{hayashi03, shinkai09}. Here we demonstrate a scalable approach to generating charge coherence by applying non-adiabatic voltage pulses to the surface depletion gates. State readout is performed using a non-invasive quantum point contact (QPC) charge detector and the DQD contains just a single electron \cite{field93}. The coherence time is extracted as a function of detuning and approaches $\sim$ 7 ns at the charge degeneracy point, where the qubit is first-order-insensitive to charge fluctuations. Comparing the data with a simple decoherence model and simulations that incorporate $\alpha/f$ noise allows us to extract the magnitude of the noise, $\alpha \sim \left(2 \times 10^{-4} e\right)^2$.

A scanning electron microscope image of a device similar to the one measured is shown in Fig.\ 1(a). The gate electrodes are arranged in a triple quantum dot geometry and deplete the two-dimensional electron gas supported by the GaAs/AlGaAs heterostructure \cite{petta10}. We form a DQD using the left and middle dots of the structure for this experiment, while the right side of the device is configured as a QPC charge detector with conductance $g_Q$. The DQD was cooled in a dilution refrigerator with an electron temperature of $\sim$80 mK and operated near the  $(1,0)-(0,1)$ charge transition, where $(n_L, n_R)$ denote the absolute number of electrons in the left and right dots.

In the one electron regime the single-particle level spacing is on the order of 1 meV and the DQD is well approximated by the two-level Hamiltonian,
\begin{eqnarray}
H =\frac{1}{2}\epsilon\sigma_z + \Delta \sigma_x,
\end{eqnarray}
with the basis states $|L\rangle$=(1,0) and $|R\rangle$=(0,1). The level detuning, $\epsilon$, is adjusted by sweeping across the interdot charge transition, as indicated in the charge stability diagram shown in Fig.\ 1(b). We adjust the interdot tunnel coupling $\Delta$ using the voltage $V_M$ on gate M. The energy splitting between the two eigenstates is given by  $\Omega(\epsilon)=\sqrt{\epsilon^2 + (2\Delta)^2}$, with a $2\Delta$ tunnel splitting at $\epsilon$=0.

We first characterize the two-level system using microwave spectroscopy, which allows us to make a direct comparison between the measured energy splitting $\Omega(\epsilon)$ and the qubit Larmor precession frequency \cite{oosterkamp98}. The application of microwaves drives transitions between the qubit ground and excited states when the energy splitting matches the photon energy. Microwave-induced charge state repopulation is directly observed using the QPC charge detector (see Fig.\ 1(c) inset) \cite{petta04}. For a continuously driven qubit, the peak height follows $h = \frac{1}{2}(1-(\sigma_{\epsilon}^{min}/\sigma_{\epsilon})^2)$, where $\sigma_{\epsilon}$ is the linewidth. Fitting the data to this form, we extract a minimum linewidth of $\sigma_\epsilon^{min} \approx 3.7$ $\mu$eV which gives a direct measure of the inhomogeneous dephasing time, $T_2^* = \sqrt{2}\hbar/\sigma_{\epsilon}^{min} \approx 250$ ps \cite{abragam}. Measurements of the peak positions as a function of microwave frequency are used to determine the interdot tunnel coupling $\Delta$. In Fig.\ 1(d) we map the resonance position as a function of microwave frequency for several values of $V_M$. For each value of $V_M$, the data are fit to the expression for $\Omega(\epsilon)$ using the `lever arm' (conversion between gate voltage and energy) and $\Delta$ as free parameters. For this device, an adjustment of $V_M$ by 20 mV results in a factor of two change in $\Delta$.

A detailed understanding of the charge noise environment can be obtained by analyzing the decay of coherent charge oscillations in the time-domain for different values of $\epsilon$. We apply a train of voltage pulses to the left gate to coherently control the charge qubit, as shown schematically in Fig.\ 2(a) \cite{Nakamura_Nature_1999}. Starting with the qubit initialized at $\epsilon \gg 2\Delta$ in state $|R\rangle$, we apply a non-adiabatic pulse with maximum detuning $\epsilon_p$ and width $t_p$. With $\epsilon_p \sim 0$, the initially prepared state $|R\rangle$ is no longer an eigenstate and evolves according to a $\sigma_x$ rotation on the Bloch sphere [Fig.\ 2(b)]. Following the pulse, the DQD returns to large positive detuning where the charge state of the qubit is read out using the QPC charge detector. The average charge state probability is acquired over $\sim10^6$ repetitions of the pulse sequence with the bulk of each cycle spent measuring the charge state. The repetition rate is comparable to the charge relaxation time $T_1 \sim 10$ ns \cite{petta04}.

\begin{figure}
\begin{center}
		\includegraphics[scale=1]{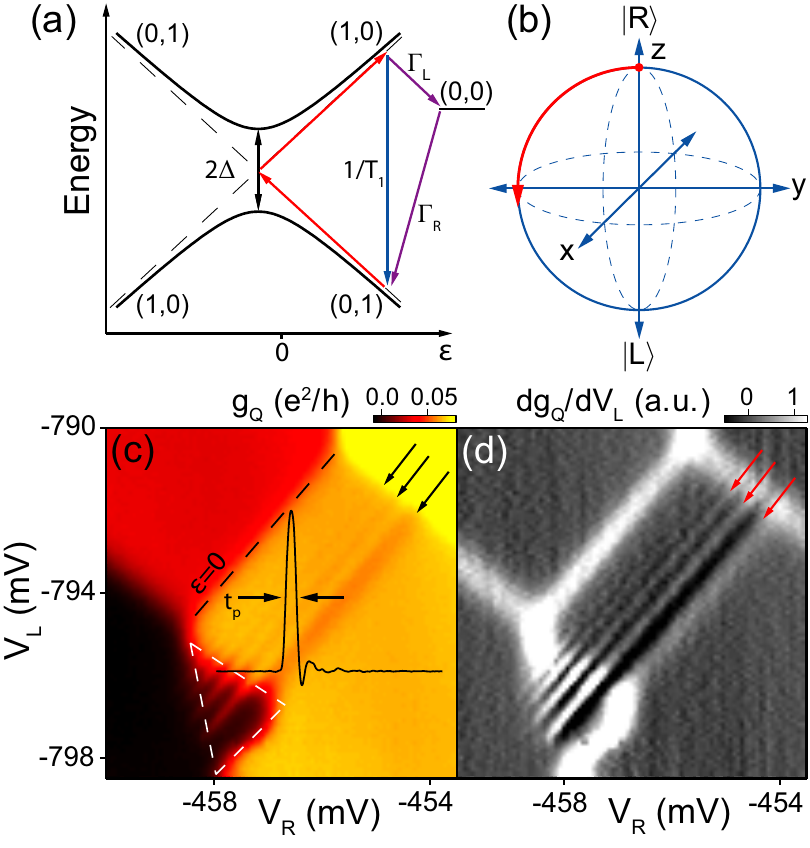}
\caption{\label{fig2} (Color online) (a) Charge qubit energy level diagram. The arrows indicate the pulse sequence and two different reset mechanisms. (b) Bloch sphere representation of charge qubit evolution at zero detuning. (c) Charge stability diagram measured with a $t_p$=150 ps pulse applied. Overlaid is a typical 150 ps pulse, as produced by the pulse generator. (d) Differentiated version of the data in (c). }
\end{center}	
\vspace{-1.0 cm}
\end{figure}

Coherent evolution of the charge qubit can be directly detected using the QPC charge detector. Figure 2(c) shows the charge stability diagram with a 150 ps pulse applied at a repetition rate of 40 MHz. In contrast with Fig.\ 1(b), we observe several resonances in the detector signal due to qubit evolution. These resonances can be more easily seen in the differentiated data shown in Fig.\ 2(d). We also observe an enhancement in the detector signal within the dashed region shown in Fig.\ 2(c), which is due to a second relaxation pathway via the (0,0) charge state, as depicted in Fig.\ 2(a).

In Fig.\ 3(a) we map the evolution of the charge state as a function of pulse width $t_p$ and pulse detuning $\epsilon_p$, which we vary by sweeping $V_L$. At large negative detunings, the oscillations rapidly decay on a timescale $\sim T_2^*$. However around $\epsilon_p = 0$ we observe a strong enhancement in the coherence time as the DQD is largely insensitive to charge noise. To further understand our data, we simulate the time evolution of the qubit using pulse profiles acquired at the output port of our pulse generator (see overlay in Fig.\ 2(c)). We account for charge noise by convolving each vertical sweep with a Gaussian with $\sigma_{\epsilon} = 3.7$ $\mu$eV width. The resulting simulations are displayed in Fig.\ 3(b) and are in good agreement with the data. In particular, the simulation reproduces the asymmetry in the data about $\epsilon_p=0$, where coherent oscillations are strongly suppressed at positive detuning due to the finite risetime of the pulses. In contrast to the simulations, the data suggest a reduced pulse amplitude at short times, which is most likely due to frequency dependent attenuation at the sample holder.

Figure 3(c) shows coherent oscillations measured at the charge degeneracy point, $\epsilon_p$=0, for two different tunnel splittings, 2$\Delta$. The observed precession rates $\nu$ agree well with the tunnel splittings extracted using microwave spectroscopy. We find that the oscillation visibility is sensitive to the experimental conditions. Detuning sweeps located near the (0,0) charge state have higher contrast, most likely due to the additional relaxation pathway. For the case of stronger tunnel couplings, $2\Delta/h = 6.6$ GHz, we observed a significant reduction in the visibility of the coherent oscillations (not shown), which the simulations suggest is due to an increase in the adiabaticity of the pulses \cite{duty04}.

\begin{figure}
\begin{center}
		\includegraphics[scale=1]{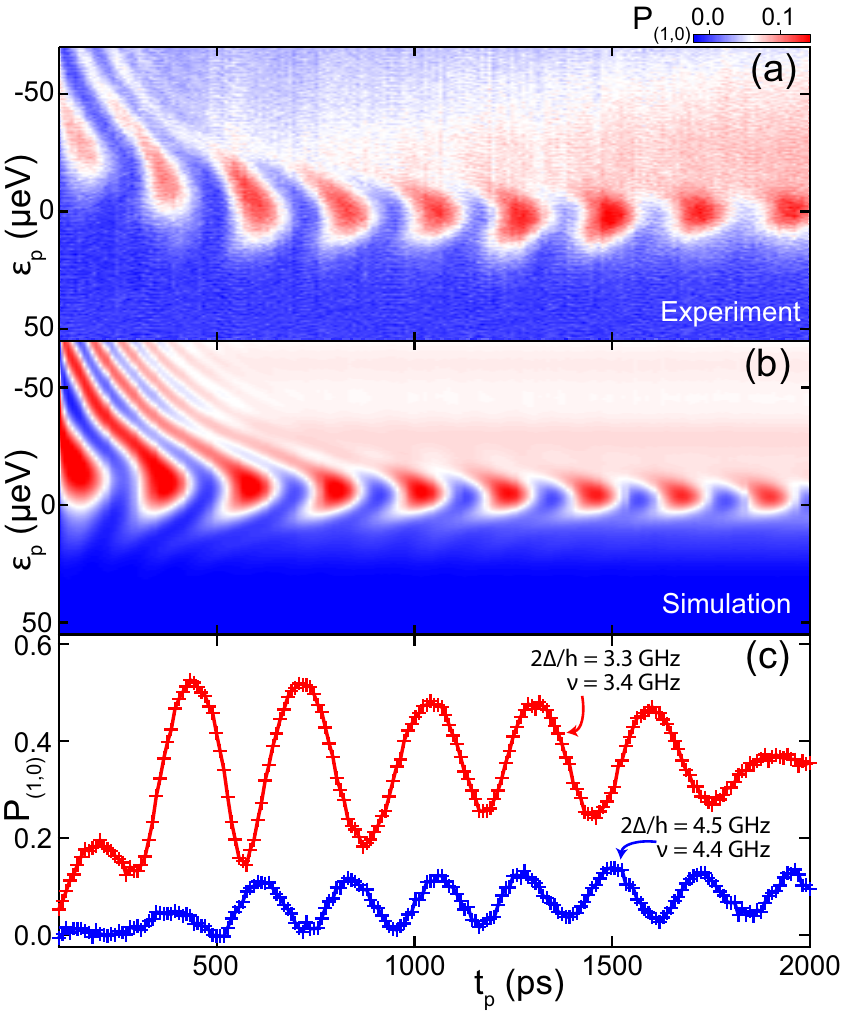}
\caption{\label{fig3} (Color online) (a) Coherent charge oscillations as a function of pulse width and detuning. (b) Simulated coherent oscillations. (c) Qubit evolution acquired at $\epsilon_p$=0 for two values of the tunnel coupling.}
\end{center}	
\end{figure}

We model the qubit coherence for a range of detunings using the coherence factor, considering only first-order coupling between the noise source and the qubit \cite{ithier05}. For a time interval $t_p$, the decay law for free Larmor precession is given by:
\begin{eqnarray}
f\left(t_p\right) =  \textrm{exp}\left[-\left(\frac{\eta}{\hbar}\right)^2 \int_{\omega_0}^\infty   S\left(\omega\right) \frac{\textrm{sin}^2 \left( t_p \omega \right/2)}{\left(\omega/2\right)^2} d \omega\right].
\end{eqnarray}
Here $S\left(\omega\right)$ is the spectral density function describing the charge fluctuations and the lower limit of the integral $\omega_0 = 2 \pi /\tau_M$ is set by the time constant of the QPC lock-in amplifier, $\tau_M \sim 100$ ms. We only consider charge noise in the detuning parameter, since detuning has a much stronger gate voltage dependance than the interdot tunnel coupling, giving $\eta = \frac{d\Omega}{d\epsilon}$.

Noise in quantum dot devices is dominated by low frequency charge fluctuators, which as an ensemble have a $1/f$ spectral density \cite{buizert08}. We therefore consider the quasi-static regime where decoherence is dominated by low-frequency Gaussian noise with a high-frequency cut-off $\omega_c$ such that $\omega_c t_p \ll 1$. In this limit, to within logarithmic factors weakly dependent on the limits of integration, the dephasing factor reduces to
\begin{eqnarray}
f\left(t_p\right) \approx \textrm{exp} \left( -\frac{1}{2} \left(\frac{\eta \sigma_{\epsilon} t_p}{\hbar}\right)^2\right),
\end{eqnarray}
where $\sigma_{\epsilon}=\sqrt{2\int_{\omega_0}^{\omega_c} S(\omega) d \omega}$ is the root-mean-square amplitude of the noise. With the coherence time defined by $f\left(T_2\right) = \textrm{exp}\left(-1\right)$, we have, $T_2 = \sqrt{2}\hbar/\eta \sigma_{\epsilon}$. The optimal operating point is at $\epsilon_p = 0$ where the energy bands are flat ($\eta = 0$) and the charge qubit is first order insensitive to charge noise \cite{vion02}.

As seen in the data in Fig.\ 3(c), the first Larmor period of coherent evolution is not captured by the traces through $\epsilon_p=0$ due to reduced pulse amplitudes at short pulse lengths. In order to make a quantitative comparison between the data and theory, we first correct the data by fitting to find the detuning value that corresponds to the tip of the voltage pulse reaching the charge degeneracy point. We then shift each trace along the detuning axis in order to align the oscillations along $\epsilon_p$=0, resulting in the corrected data shown in Fig.\ 4(a). In Fig.\ 4(b) we plot the coherent oscillations extracted from the corrected data set at three different values of $\epsilon_p$, as indicated in Fig.\ 4(a). We fit the oscillations to a damped cosine form, $a t_p + b\times\textrm{exp}\left( -(t_p/T_2)^2 \right) \textrm{cos}\left(t_p/c + d \right)$ with $a$, $b$, $c$, $d$ and $T_2$ as free parameters. The linear coefficient $a \sim 0.02 $ ns$^{-1}$ accounts for a small upward drift in the charge occupancy, which is presently not well understood.

\begin{figure}[t]
\begin{center}
		\includegraphics[scale=1]{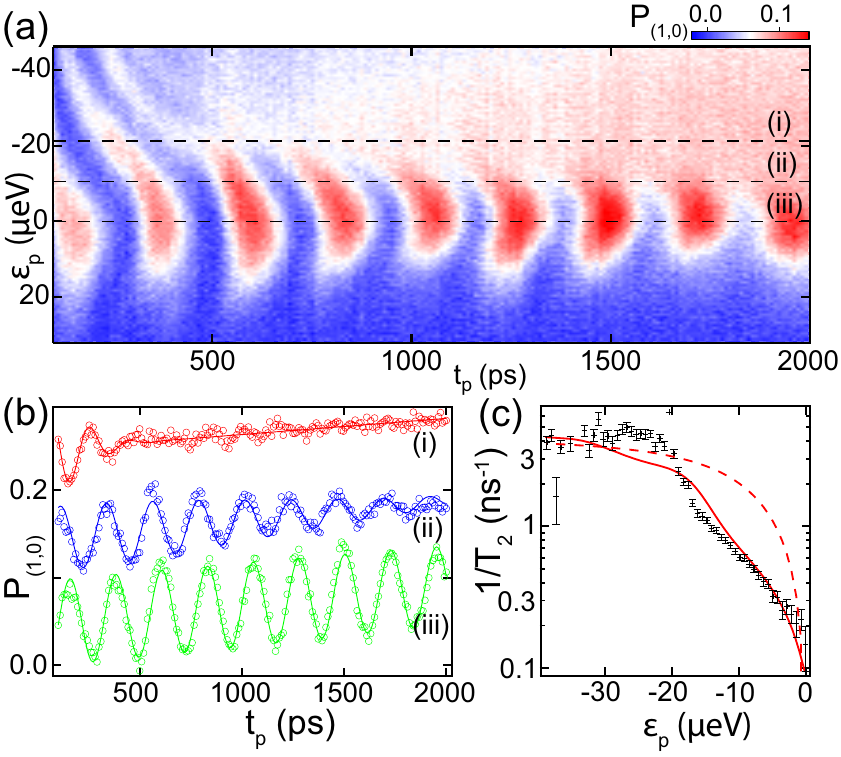}
\caption{\label{fig4} (Color online) (a) Data from Fig.\ 3(a) corrected to take into account the reduced pulse amplitude at short pulse widths. (b) Fits to the coherent oscillations at three different detunings as indicated in (a). (c) Coherence rates extracted from the fits as a function of detuning. The dashed line shows the expected behavior based on a simple low frequency noise model with $\sigma_\epsilon = 3.7$ $\mu$eV. The solid line shows the coherence rates extracted from simulations with $\sigma_\epsilon =  5$ $\mu$eV.}
\end{center}	
\vspace{-0.5 cm}
\end{figure}

The extracted coherence rate, $1/T_2$, is plotted as a function of $\epsilon_p$ in Fig.\ 4(c). The dashed curve in Fig.\ 4(c) is a fit to equation 3 with a best-fit $\sigma_\epsilon = 3.9$ $\mu$eV, consistent with the photon assisted tunneling peak widths. Assuming $1/f$ noise with a spectral density function $S(\omega) = \left(\frac{E_c}{e}\right)^2 \frac{\alpha}{|\omega|}$, where $E_c \sim 3.2$ meV is the charging energy for one of the quantum dots and taking $\omega_c/2\pi = 40$ MHz, we estimate $\alpha \sim \left(2 \times 10^{-4} e\right)^2$. At large negative values of detuning ($\epsilon_p < -20$ $\mu$eV) theory is in reasonable agreement with the data. However, approaching zero detuning we observe deviations from the simple dephasing model. The solid curve in Fig.\ 4(c) shows coherence rates extracted from the simulated data with $\sigma_{\epsilon} = 5$ $\mu$eV. This curve provides a better fit to the experimental data and suggests that higher order coupling terms not taken into account in the simple model (Eq.\ 3) are significant.

The coherence at zero detuning may become limited by other mechanisms. Fits to our data give $T_2 =  7\pm 2.5$ ns. At longer pulse lengths the oscillations periodically decay and then re-emerge, making it difficult to more accurately determine the coherence time \cite{epaps}. The $\sim$ 3.5 ns period of the beating approximately matches the time taken for a signal to travel back and forth between the sample and the bias-tee and has been observed in other experimental setups \cite{guillaume04}. Nonetheless, the decay envelope of the beating also suggests a coherence time of $T_2 \sim 10$ ns. The coherence time is of the same order as typical charge relaxation times in GaAs DQDs, suggesting that this could be the limiting mechanism ($T_2 \leq 2 T_1$). The observed coherence time is longer than the previously reported value of $\sim 1$ ns, which was limited by the strong tunnel coupling to leads required for the readout process \cite{hayashi03}.

Finally, we consider how charge noise might impact spin qubit operation. In the case of a spin $\sqrt{\textrm{SWAP}}$ operation involving the effective exchange energy $J\left(\epsilon\right) \approx \frac{1}{2}\left(\epsilon + \sqrt{\epsilon^2 +\Delta^2}\right)$, the gate error probability (at $\epsilon \ll -\Delta$) is given by,
\begin{eqnarray}
\frac{t_{op}}{T_2} \approx \left(\frac{\pi}{2}\right)^2 \frac{\sigma_{\epsilon}\hbar}{\sqrt{2}t_{op}\Delta^2},
\end{eqnarray}
where $t_{op} = \frac{\pi}{2}\hbar/J\left(\epsilon\right)$ is the gate operation time. Taking a typical gate operation time $t_{op} = 1$ ns with $\Delta = 8$ $\mu$eV and $\sigma_{\epsilon} = 4$ $\mu$eV gives an error probability of $\sim7$\%.

\begin{acknowledgments}
Research at Princeton was supported by the Sloan and Packard Foundations, DARPA grant N66001-09-1-2020, and the NSF through the Princeton Center for Complex Materials (DMR-0819860) and CAREER award (DMR-0846341). Work at UCSB was supported by the DARPA and the UCSB National Science Foundation DMR Materials Research Science and Engineering Center.
\end{acknowledgments}


\end{document}